\documentstyle[12pt,aaspp4,tighten]{article}
\begin{document}

\newcommand{\msol}{M{$_{\odot}$}} 
\newcommand{\lsol}{L{$_{\odot}$}}
\newcommand{\rsol}{R{$_{\odot}$}}
\newcommand{\kms}{km~s{$^{-1}$}} 
\newcommand{\cmq}{cm{$^{-2}$}} 
\newcommand{\cmc}{cm{$^{-3}$}} 
\newcommand{\hii}{H{\sc ii}}
\newcommand{\hcop}{HCO$^+$($1\rightarrow 0$)}
\newcommand{\hicop}{H$^{13}$CO$^+$($1\rightarrow 0$)}
\newcommand{\hicn}{H$^{13}$CN($1\rightarrow 0$)}
\newcommand{\sio}{SiO(v=0, $2\rightarrow 1$)}
\newcommand{\ho}{H$_2$O}
\newcommand{\degr}{$^{\circ}$}
\newcommand{\asec}{$^{\prime \prime}$}                   
\newcommand{\um}{$\mu$m}                                 
\newcommand{\mdot}{\.{M}}
\newcommand{\msunyr}{$M_{\odot}$ yrs$^{-1}$}
\newcommand{\lsim}{\;\lower.6ex\hbox{$\sim$}\kern-7.75pt\raise.65ex\hbox{$<$}\;}
\newcommand{\gsim}{\;\lower.6ex\hbox{$\sim$}\kern-7.75pt\raise.65ex\hbox{$>$}\;}
\newcommand{\hi}{{\it High}}
\newcommand{\lo}{{\it Low}}

\lefthead{Molinari, Testi, Brand et al.}
\righthead{A Prototype Massive Class 0 Object}

\title{IRAS\,23385+6053: A Prototype Massive Class 0 Object\footnotemark[0]}

\author{Sergio Molinari \altaffilmark{1}, 
   Leonardo Testi \altaffilmark{2}, 
   Jan Brand \altaffilmark{3},
   Riccardo Cesaroni \altaffilmark{4} and 
   Francesco Palla \altaffilmark{4}}

\footnotetext[0]{Based on observations with ISO, an ESA project with
instruments funded by ESA Member States (especially the PI countries:
France, Germany, the Netherlands and the United Kingdom) with the
participation of ISAS and NASA.}
\altaffiltext{1}{Infrared Processing and Analysis Center, California
Institute of Technology, MS~100-22, Pasadena, CA~91125
(molinari@ipac.caltech.edu)}
\altaffiltext{2}{Division of Physics, Mathematics and Astronomy,
California Institute of Technology, MS~105-24, Pasadena, CA~91125
(lt@astro.caltech.edu)}
\altaffiltext{3}{Istituto di Radioastronomia - CNR, Via Gobetti 101,
I-40129, Bologna (brand@astbo1.bo.cnr.it)}
\altaffiltext{4}{Osservatorio Astrofisico di Arcetri, Largo E. Fermi 5,
I-50125, Firenze (palla, cesa@arcetri.astro.it)}

\begin{abstract}

IRAS\,23385+6053 is a Young Stellar Object with luminosity $\sim
1.6\times 10^4$ \lsol\ at a kinematic distance of 4.9~kpc. This candidate
precursor of an ultracompact \hii\ region is associated with a
millimeter source detected at the JCMT but is undetected at centimeter
wavelengths with the VLA. We observed this source with the OVRO 
millimeter array at 3.4~mm in the continuum, \hcop, \hicop\ and \sio\ line 
emission, and with CAM aboard ISO at 6.75\um\ and 15\um.
The IRAS source is coincident with a 3.4~mm compact
($r_{core}\simeq0.048$~pc) and massive ($M\simeq370$~\msol) core, which
is undetected at 15\um\ to a 3$\sigma$ level of 6~mJy; this is
compatible with the derived H$_2$ column density of $\sim 2\times
10^{24}$\cmq\ and the estimated visual extinction A$_V\sim$2000 mag.
We find L$_{submm}$/L$_{bol}\sim3\times10^{-3}$ and
M$_{env}$/M$_{\star}\gg 1$, typical of Class 0 objects.  The source is
also associated with a compact outflow characterized by a size
$\lsim$r$_{core}$, a dynamical timescale of $\lsim 7 \times
10^3$~years, and a mass loss rate $\dot{M}\gsim10^{-3}$\msol\,yr$^{-1}$. The
axis of the outflow is oriented nearly perpendicular to the
plane of the sky, ruling out the possibility that the non-detection at
15~\um\ is the result of a geometric effect. 
All these properties suggest that IRAS\,23385$+$6053 is the
first example of a {\it bona fide} massive Class 0 object.

\end{abstract}


\setcounter{footnote}{0}
\section{Introduction}
\label{introduction}

The last few years have seen a rapid growth in observations aimed at
identifying intermediate- and high-mass star forming sites in a wide
range of evolutionary stages ranging from ``Hot Cores'' (Cesaroni et
al.~\cite{Cetal94a}), to ultracompact \hii\ (UC\hii) regions (Wood \&
Churchwell~\cite{WC89}), to proto-Ae/Be stars (Hunter et
al.~\cite{Hetal98}). The characterization of the earliest stages of
high mass-star formation is more difficult than for low-mass objects,
given their shorter evolutionary timescales. The likely candidates must
be luminous, embedded in dense circumstellar environments, and
non-associated with \hii\ regions (c.f. Habing \& Israel~\cite{HI79})

We have undertaken a systematic study to identify a sample of massive
protostellar candidates, selecting sources from the IRAS-PSC2 catalogue
and filtering them according to the above criteria. On the basis of
\ho\ maser (Palla et al.~\cite{Petal91}), ammonia lines (Molinari et
al.~\cite{Metal96}), centimeter and submillimeter continuum
observations (Molinari et al. \cite{Metal98a}; \cite{Metal98b}), we
were finally left with about a dozen candidates which may be precursors
of UC\hii\ regions. Several of these sources have been recently
observed with the IRAM-30m telescope (Brand et al. 1998, in
preparation) and some with the Infrared Space Observatory (ISO), to
characterize them over a wide range of wavelengths. We are now
beginning a program of high spatial resolution millimeter wave
observations of these objects. This {\it Letter} presents the results
for the first source imaged, IRAS\,23385+6053, at a kinematic distance
\footnote{In Molinari et al. (1996), we quoted a distance of 6.9 kpc
derived from the {\it observed} galactic velocity field, which should
however not be used for d$\geq$5 kpc in the 2$^{nd}$ and 3$^{rd}$
quadrants.} of 4.9~kpc (using the Brand \& Blitz~(\cite{BB93}) galactic
rotation curve).  According to the definition (e.g.
Andr\'e~\cite{A96}), this is an excellent candidate of a high-mass
counterpart of Class 0 objects.  It is associated with an \ho\ maser
and with NH$_3$ line and submillimeter continuum emission, while no
centimeter continuum emission was detected in our VLA observations down
to a level of $\sim 0.5$~mJy/beam.

\section{Observations}
\label{obser}


The field centered on IRAS\,23385$+$6053 was observed at 88~GHz
(continuum and lines) with the Owens Valley Radio Observatory (OVRO)
millimeter wave array during January/February 1998. Two configurations
of the six 10.4-m telescope antennas provided baselines in the range
15--240~m. Cryogenically cooled SIS receivers have typical average
system temperature of $\sim 350$~K.  The continuum observations used
two 1~GHz wide bands of an analog correlator.  For the line
observations, the digital correlator was split into several bands,
enabling simultaneous observations of the \hcop\ and \hicop\ rotational
lines with $\sim 20\rm\,\, km\,s^{-1}$ bandwidth and $0.84\rm\,\,
km\,s^{-1}$ resolution; the \hcop\ line was also observed with a $\sim
80\rm\,\, km\,s^{-1}$ bandwidth and $3.4\rm\,\, km\,s^{-1}$
resolution.  Additionally, data on the \sio\ line were acquired in the
image sideband of the \hcop\ line with the same bandwidths and
resolutions.  Gain and phase were calibrated by means of frequent
observations of the quasar 0224$+$671.  3C273 was used for passband
calibration. The flux density scale was determined by observations of
planets and the estimated uncertainty is less than 20\%.  The
synthesized beamsize is $4\farcs 1 \times 3\farcs 5$. Noise levels in
the maps are $\sim 0.7\rm\,\, mJy/beam$, $\sim 20\rm\,\, mJy/beam$, and
$\sim 45\rm\,\, mJy/beam$ for the continuum and the low and high
resolution spectral data, respectively.  


ISO (Kessler et al.~\cite{Ketal96}) observations with CAM (Cesarsky et
al.~\cite{Cetal96}) were made in staring mode on June 18 1997.
Diffraction-limited images (resolution 2\asec.2 and 5\asec,
respectively) were obtained in the two broad-band filters LW2
($\lambda_{eff}$=6.75\um, $\Delta \lambda$=3.5\um) and LW3
($\lambda_{eff}$=15.0\um, $\Delta \lambda$=6\um). The
pixel-field-of-view is 3\asec, and the absolute pointing accuracy of
the satellite was presumably within $\sim3$\asec\ at the time of the
observations. More than 50 frames per filter were obtained with a 2s
integration time per frame, and the expected 3$\sigma$ sensitivity
limits were 0.33 and 0.45 mJy/arcsec$^2$ in LW2 and LW3, respectively.
The data have been reduced using CAM Interactive Analysis (V2.0), and
the absolute calibration is accurate within 20\%.

\section{Results}

\begin{figure}
\plotfiddle{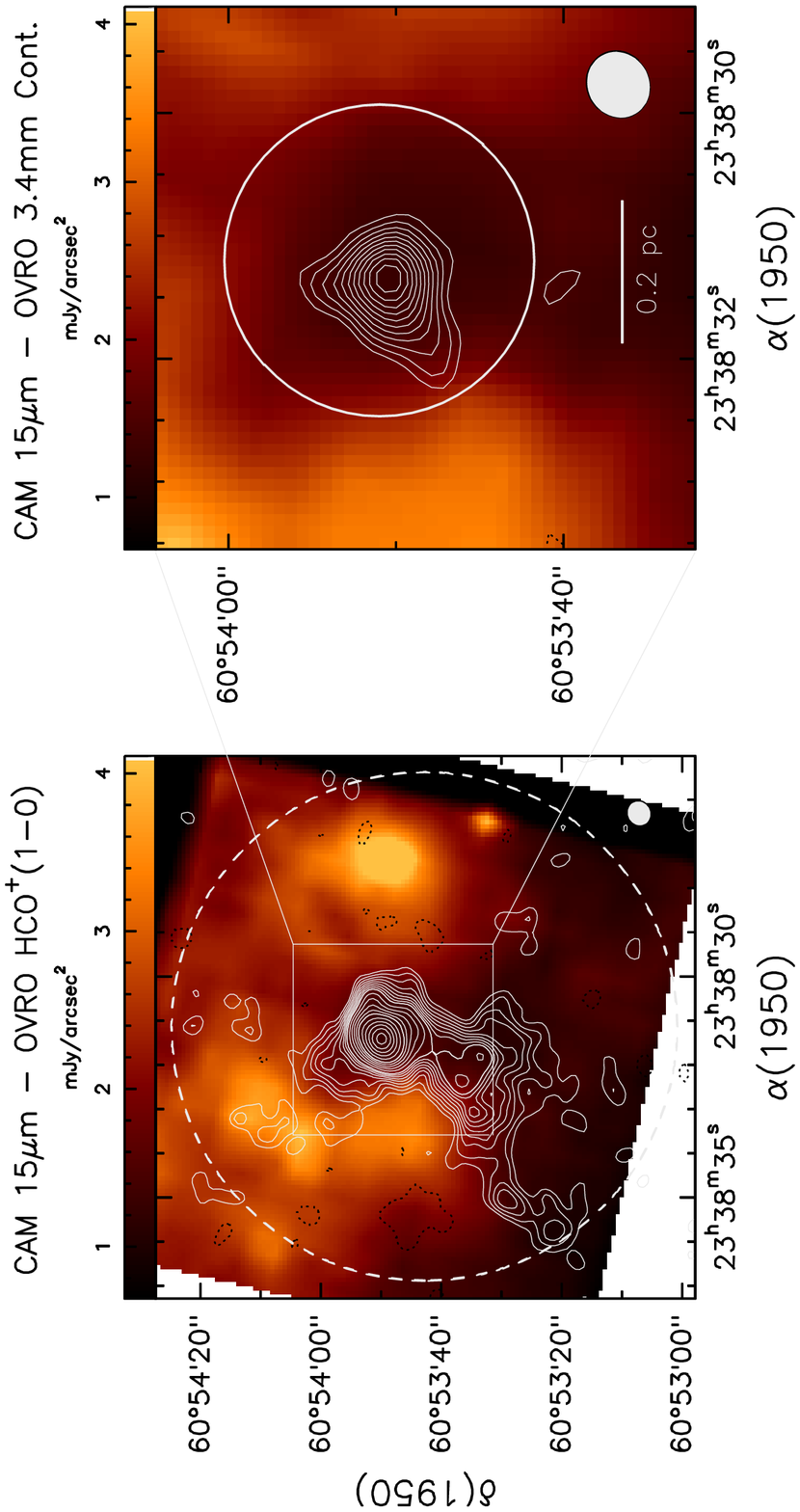}{5.5in}{-90}{80}{80}{-300}{400}
\caption[plates.ps]{\label{plates}
a) CAM LW3 filter (15\um) image of IRAS~23385+6053, with superposed the
contour plot of \hcop\ line intensity integrated over a velocity
interval of $\sim7$~\kms\ centered at V$_{LSR}=-51.0$~\kms; the CAM
image has been resampled at the same pixel size used to reconstruct the
OVRO maps. To avoid saturation at the peak emission, contours are
spaced from 0.3 to 1.5 by 0.15, from 1.5 to 3.0 by 0.3, and from 3.5 to 5.6
by 0.5~Jy~\kms/beam. The dashed circle shows the OVRO primary beam
HPBW.  - b) Enlargement of the previous image around the central
position with superposed the contour plot of the OVRO 3.4mm continuum
emission from 2.0 to 9.0 by 0.7 mJy/beam. The circle shows the JCMT
19\asec\ aperture centered at the 1.1~mm peak position. The grey
ellipse is the OVRO synthesized beam FWHP.}
\end{figure}

Figure~\ref{plates}a shows the contour plot of the
\hcop\ integrated line intensity overlaid on the CAM-LW3 filter
15~\um\ image. Figure~\ref{plates}b shows an enlargement
of the central region, with a contour plot of the 3.4~mm continuum.
With the exception of \hcop\ emission in one area to the NE, the
HCO$^+$ line and the 3.4~mm continuum arise from the central region
where the 15~$\mu{m}$ continuum is faintest; HCO$^+$ also shows an
extension towards S-SE.

In the dark region at the center of the 15~\um\ image, flux densities
of the order of $\sim$1~mJy/arcsec$^2$ are found. Similar values are
also found in the southern portion of the CAM field, suggesting that
such a constant level is due to faint foreground diffuse emission.  The
$rms$ noise computed in the southern 20\asec\ of the image is
$\sim$0.15~mJy/arcsec$^2$. Similar flux densities and $rms$ levels are
obtained for the 6.75~\um\ image. We thus conclude that no
15~\um\ source is detected at the position of the central 3.4~mm
continuum peak down to a $3\sigma$ level of $\sim 0.45$~mJy/arcsec$^2$
which, integrated over the deconvolved area of the 3.4~mm core,
corresponds to $\sim6$ mJy.

The 3.4~mm continuum core is compact with slight elongations to the SE
and N (Fig.~\ref{plates}). The total integrated flux density is 19~mJy
and the deconvolved source size is 4\asec.5$\times$3\asec.6,
corresponding to an average radius of 0.048~pc at a distance of
4.9~kpc. The continuum emission peaks at  $\alpha(1950)=23^{\rm h}38^{\rm
m}31^{\rm s}.4,~\delta(1950)=+60^{\circ}53^{\prime}50$\asec, and is
probably due solely to dust thermal emission since the extrapolation at
3.4~mm of the~2 cm upper limit with an ionized wind model $\sim\nu^{0.6}$
(Panagia \& Felli~\cite{PF75})
yields less than 10\% of the observed flux density.

\begin{figure}
\plotone{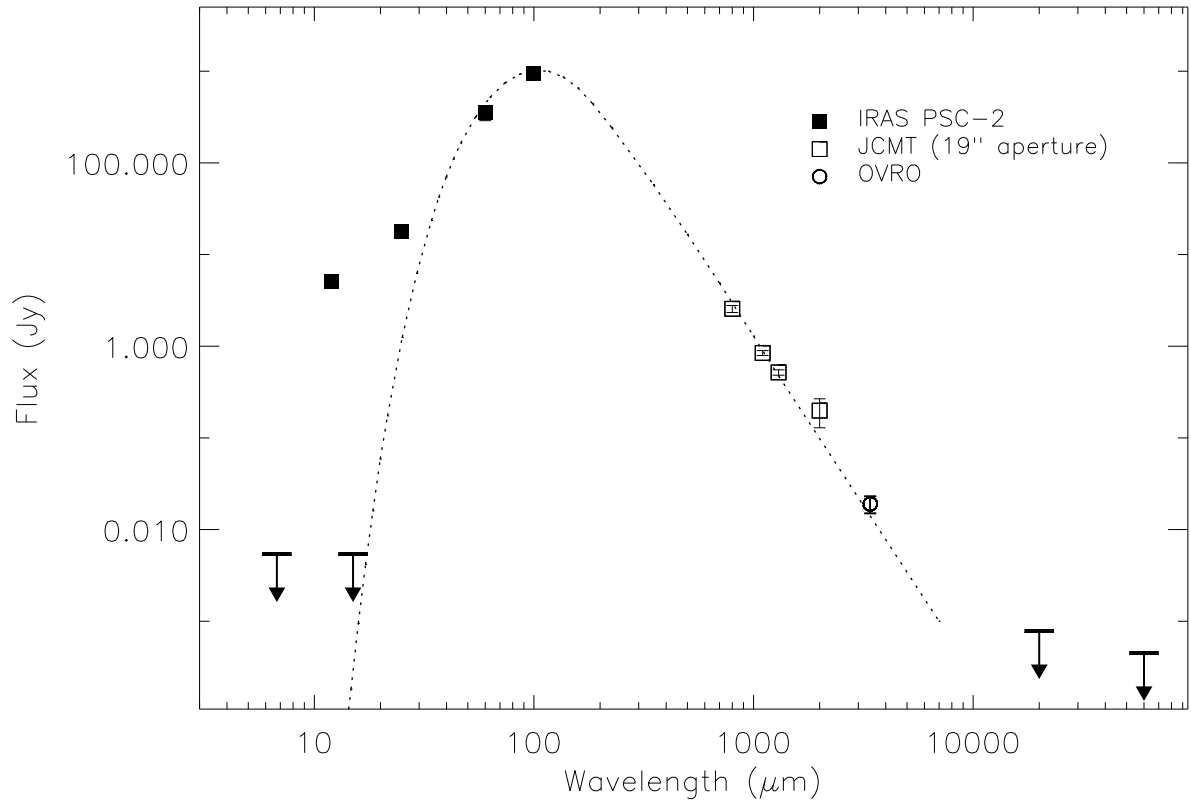}
\caption[fsed.ps]{\label{sed}
Spectral energy distribution of the IRAS~23385$+$6053 core. 
The dotted line shows the model fit described in the text.}
\end{figure}

Fig.~\ref{sed} shows the complete spectral energy distribution (SED) of
IRAS\,23385+6053.  The OVRO flux density is consistent with those from
JCMT (Molinari et al.~\cite{Metal98b}), indicating that no significant
diffuse millimeter emission has been missed by the interferometer. On
the other hand the IRAS flux densities include contributions of all the
sources seen in the 15~\um\ image of Fig.~\ref{plates}. Indeed, the
IRAS 12~\um\ and 25~\um\ points in Fig.~\ref{sed} (5.05 and 17.8 Jy,
respectively) are compatible with the CAM 6.75 and 15 \um\ flux integrated
over all the bright emission visible on the CAM image ($\sim$4.8 and
9.3 Jy respectively, in an area of 6.8$\times10^{-8}$
sterad), while they lie more than 3 orders of magnitude above the CAM flux
density limits for the core estimated above.
Clearly, the 6.75 and 15~\um\ emission area, which is not directly
related to the millimeter core, is responsible for the IRAS 12 and
25\um\ flux densities; however, it is unlikely that it contributes for
a significant fraction of the 60~\um\ and 100~\um\ IRAS flux densities 
because:
1) at millimeter wavelengths, the core is compact and no discrete or
diffuse continuum emission is detected at either OVRO or the JCMT;
2) a power law extrapolation of the 6.75, 12, 15 and 25 \um\ flux
densities to longer wavelengths, leads to flux densities at 60 and
100~\um\ $\lsim$ 10\% of the observed values. We conclude that the
millimeter core is the only source contributing significantly at 60 and
100~\um. 

In Fig.~\ref{sed} the dotted line is a fit to the data, obtained
adopting a spherical envelope model where density and temperature vary
according to radial power laws with exponents -0.5 and -0.4
respectively.  The density and
temperature at the external radius, which is the observed
r$_{core}\simeq0.048$pc, are 7$\times10^6$\cmc\ and 40 K. The opacity is
$\kappa_\nu=\kappa_{1.3mm}\,(\lambda({\rm mm})/1.3)^{-1.9}$,
$\kappa_{1.3mm}=0.005\rm\,\,cm^2\,g^{-1}$
(Preibisch et al.~\cite{Pea93}).  We compute the bolometric luminosity
of the source by integrating the data with a power law interpolation
scheme, excluding the IRAS 12 and 25~\um\ flux densities and using the
CAM upper limit at 15~\um. At the adopted distance of 4.9~kpc, we find
L$_{bol}\simeq 1.6\times10^4$~\lsol.  Integrating the fitted density
power law and adopting a gas to dust ratio of 100 by mass, we estimate
a mass of $\sim$ 370~\msol\ and a mean H$_2$ column density of $\sim
2\times10^{24}$~cm$^{-2}$ for the core (corresponding to a visual
extinction A$_{\rm V}\sim 2000$).  We also computed the mass of the
core from HCO$^+$ line emission. The optical depth in each channel
within 3 \kms\ of V$_{LSR}$=$-$51 \kms\ was estimated from the
HCO$^+$/H$^{13}$CO$^+$ ratio, assuming a $^{12}$C/$^{13}$C$=87$
abundance ratio using the galactic abundance gradient and the local-ISM
value given in Wilson \& Rood~(\cite{WR94}) and a source galactocentric
distance of $\sim 11$~kpc; we find that the core of the main isotope
line is optically thick.  We derive a lower limit (due to the optical
depth) for the H$_2$ mass of 180 \msol\ assuming T$_{ex}=30$~K, based
on the peak \hcop\ synthesized beam brightness temperature, and
[HCO$^+$]/[H$_2$]$=10^{-9}$. This value compares reasonably well with a
virial mass of 160~M$_\odot$ obtained from the line width ($\sim
3.5\rm\,\,km\,s^{-1}$) and \hcop\ core radius ($\sim 0.06$~pc).

\begin{figure}
\plotone{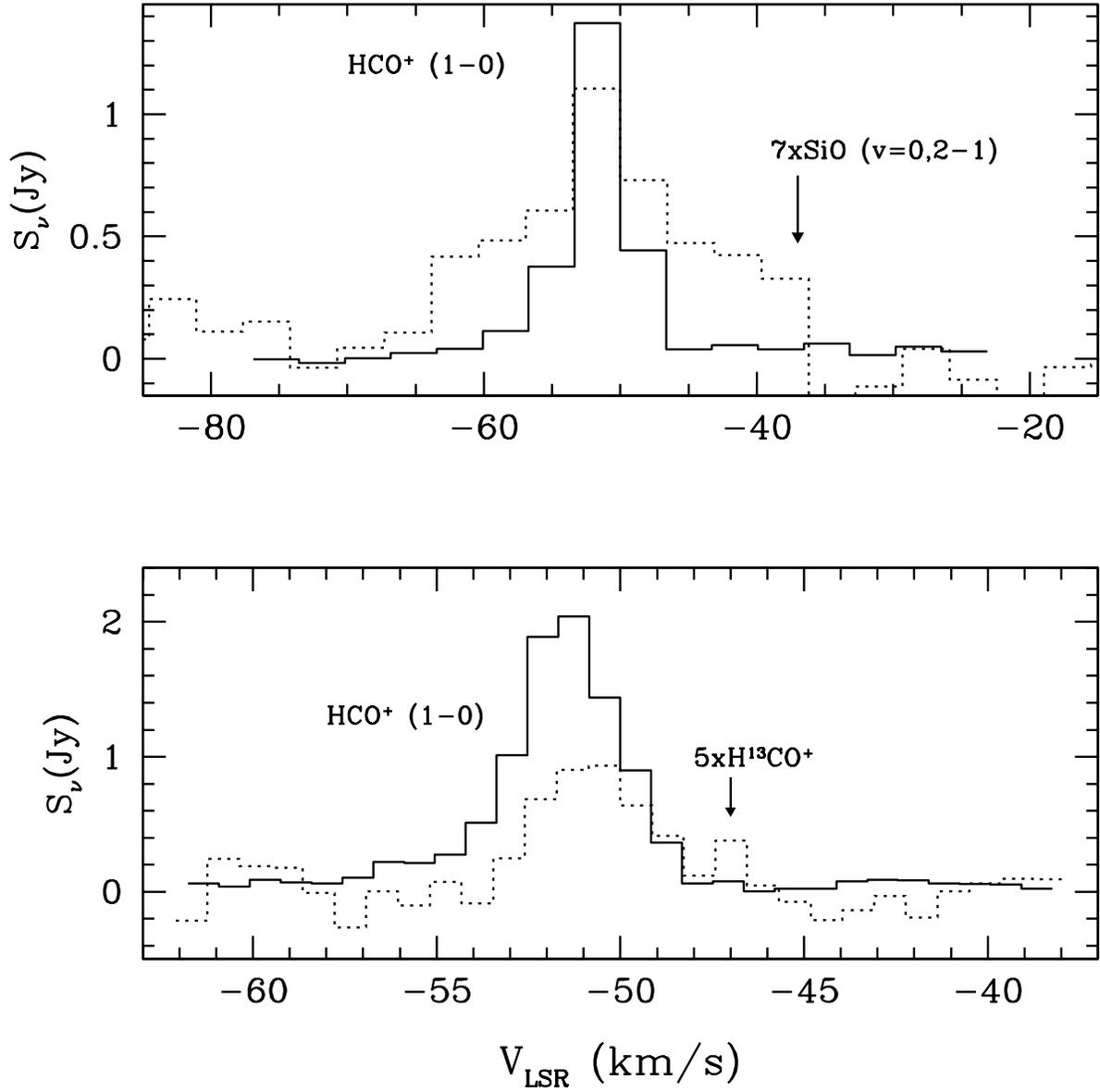}
\caption[fspec.ps]{\label{fspec}
OVRO spectra integrated over $\sim20$~arcsec$^2$ area 
centered on the core. Top panel: low resolution, {\it rms}=0.02 Jy.
Bottom panel: high resolution, {\it rms}=0.04 Jy. In all panels, the vertical
scale is the flux density in Jansky.}
\end{figure}
\begin{figure}
\plotfiddle{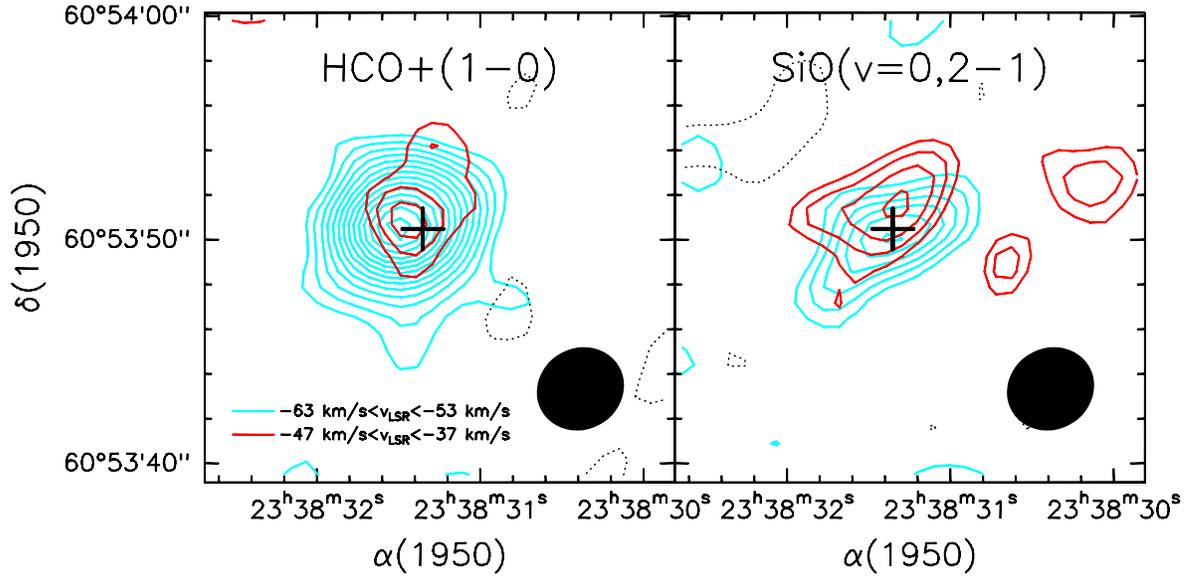}{5in}{-90}{75}{75}{-280}{400}
\caption[flobes.ps]{\label{flobes}
Outflow maps of IRAS\,23385+6053. Contour levels start at 0.3 and
increment by 0.12~Jy~\kms\ for the two maps; the dotted contour
corresponds to -0.3~Jy~\kms.  The cross represents the position of the
peak in the 3.4 mm continuum map.  The full ellipse is the OVRO beam}
\end{figure}

In Fig.~\ref{fspec} we present profiles of the \sio\ and \hcop\ lines
at low spectral resolution and the \hcop\ and \hicop\ lines at higher
resolution,  integrated over the $\sim 20$~arcsec$^2$ centered on the
core peak.  Both the SiO and the HCO$^+$ profiles show broad wings,
while no wing emission is detected from \hicop. Images of the HCO$^+$
and SiO emission integrated over the blue (from $-$63 to $-$53~\kms)
and the red (from $-$47 to $-$37~\kms) wings are shown in
Fig.~\ref{flobes}. A compact outflow centered on the continuum source
is detected; both the blue and the red lobes are visible in \sio,
whereas only the blue lobe is clearly detected in \hcop, although a
faint red lobe may be present. The lobes are barely resolved; their
position relative to each other, i.e. their degree of overlap, suggests
that the inclination of its axis can be no more than 30$^{\circ}$ with
respect to the line of sight (Cabrit \& Bertout~\cite{CB86};
\cite{CB90}). This orientation of the outflow rules out any possibility
that the large extinction derived toward IRAS\,23385+6053 and the
non-detection at 15 \um\ may result from geometrical effects (e.g. an
edge-on disk). Properties of the IRAS\,23385+6053 outflow are listed in
Table~\ref{flowpar}.  Column density, mass, momentum and energy are
computed in each velocity channel and finally summed, while the
dynamical timescale is averaged over the velocity channels.  We
estimated the column density of material assuming LTE and optically
thin approximations in all the line wings channels within the velocity
ranges listed previously, assuming the above determined excitation
temperature of 30~K for both HCO$^+$ and SiO.  For both molecules we
adopt an abundance fraction of 10$^{-9}$ relative to H$_2$. The
velocity measured in the lobes is a good estimate of the true gas
velocity because the outflow axis is nearly perpendicular to the plane
of the sky. This configuration prevents an accurate measure of lobe
size, but a reasonable estimate can be made based on the non-detection
at 15\um. Since any outflow more extended than the core would have
created a cavity through which mid-infrared radiation could have
escaped, we infer that the extent of the flow lobes is less than the
core size.  The size of the lobes and the dynamical time scale of the
outflow are therefore upper limits, while the mass loss rate and
kinetic luminosity are lower limits.

The flow mechanical power and mass loss rate in Table~\ref{flowpar} are
high compared to values characterizing outflows powered by low-mass
protostars (e.g.  Fukui et al.~\cite{Fuetal93}), but in good agreement
with those for high-mass young stellar objects (Shepherd \&
Churchwell~\cite{SC96}).  Nevertheless, the dynamical timescale is
similar to values found for the youngest outflows around low-mass Class
0 objects (Andr\'e et al.~\cite{AWT93}).

\section{IRAS\,23385+6053: A Massive Class 0 Object}
\label{discussion}

IRAS\,23385+6053 was not detected at 6 and 2~cm with the VLA B-array
with 3$\sigma$ sensitivity limits of $\sim$0.5 and $\sim$0.8 mJy/beam
respectively (Molinari et al.~\cite{Metal98a}), implying that any
associated \hii\ region, if present, must be optically thick and very
compact. In this latter case, from the VLA synthesized beamsize and
assuming that the source has not been detected because of beam
dilution, the 2~cm upper limit leads to an upper limit for the radius
of an optically thick, spherical and homogeneous \hii\ region
$r_{\hii}\lsim 100$~AU (see also Molinari et al.~\cite{Metal98a}).  An
\hii\ region originating from a B0 ZAMS star (which has a luminosity
comparable to the bolometric luminosity of IRAS\,23385+6053) in a dense
($n_{\rm H_2}\sim 10^{7}$~\cmc) environment takes only few years to
expand to a radius of 100 AU (De Pree et al.~\cite{Detal95}). However,
a modest accretion rate of $4\times10^{-6}$~\msol/yr could squelch the
expanding \hii\ region (see e.g.  Walmsley~\cite{W95}), thus explaining
the properties of our source in terms of a heavily obscured B0-ZAMS
with a residual accretion.  Another possible explanation is that there
is no \hii\ region because IRAS\,23385+6053 has not yet reached the
ZAMS. In this case we assume that the observed bolometric luminosity is
due to accretion onto a protostellar core with accretion luminosity
expressed as L$_{acc}=3.14\times10^4$ \lsol\ (M$_{\star}$/\msol)
(\rsol/R$_{\star}$) ($\dot{\rm M}/10^{-3}$\msol\,yr$^{-1}$).  Since
L$_{acc}$ is known, we can combine it with the mass-radius relation of
Stahler, Palla \& Salpeter (\cite{SPS86}), R$_{\star}=27.2\times$
\rsol\ (M$_{\star}$/\msol)$^{0.27}$ ($\dot{\rm
M}/10^{-3}$\msol\,yr$^{-1}$)$^{0.41}$, to derive the protostellar mass
M$_{\star}$ as a function of the accretion rate. For an accretion rate
$\dot{\rm M}\sim10^{-3}$ \msol\,yr$^{-1}$, which is expected if the
outflow (see Table~\ref{flowpar}) is driven by infall (e.g. Shu et al.
1988), we obtain M$_{\star}$=39 \msol, implying that the embedded IRAS
source is a massive protostar. In either case, this source is evidently
massive and extremely young.

Based on the evidence presented here, {\it we propose IRAS\,23385+6053 as a
prototype high-mass Class~0 object}: it is undetected in the mid-IR
($\lambda<15$\um) and in the radio continuum; it has
L$_{submm}$/L$_{bol}\gtrsim 5\times10^{-3}$ (where L$_{submm}$ is
obtained by integrating longward of 350\um); it has a ratio of the
envelope to stellar mass greater than 1 (see Andr\'e et
al.~\cite{AWT93}), M$_{env}$/M$_{\star}\gsim10$. The very short
dynamical timescale ($\lsim$7000 yrs) of the associated compact
molecular outflow is also compatible with values found for outflows
around low-mass Class 0 objects.

\noindent
{\bf Acknowledgements:}
We thank Anneila Sargent for a critical reading of the manuscript. 
Support from NASA's {\it Origins of Solar Systems} program (through grant
NAGW--4030) is gratefully acknowledged. The Owens Valley
millimeter-wave array is supported by NSF grant AST-96-13717. Research
at Owens Valley on young star and disk systems is also supported by the
{\it Norris Planetary Origins Project}. The ISOCAM data presented in
this paper were analysed using ``CIA'', a joint development by the ESA
Astrophysics Division and the ISOCAM Consortium. This project was
partly supported by CNR grant 97.00018.CT02 and by ASI grant ARS 96-66
to the Osservatorio di Arcetri.


%
%
\clearpage
\clearpage
\clearpage
\clearpage
%
%

\clearpage

\begin{deluxetable}{lcc}
\footnotesize
\tablecaption{Outflow Parameters \label{toutflow}\label{flowpar}}
\tablewidth{0pt}\tablehead{
\colhead{Parameter}& \colhead{\hcop} & 
\colhead{\sio}}
\startdata
N$_{\rm H_2}$(cm$^{-2}$)&7.5$\times$10$^{22}$&1.4$\times$10$^{23}$\nl
M$_{\rm H_2}$(\msol)& 10.6 & 19.6  \nl
P = $\sum_i M_i v_i$ (\msol \kms) & 66.6 & 159.6 \nl
E=${{1}\over{2}}\sum_i M_i 
v_i^2$(ergs)&4.7$\times$10$^{45}$&1.4$\times$10$^{46}$\nl
$<\tau_{dyn}>$ (yr) & 7300 & 5600\nl
\mdot(\msol\,yr$^{-1}$)&1.5$\times10^{-3}$&3.5$\times 10^{-3}$ \nl
$\rm\dot{P}$ = P/$\tau_{dyn}$ &9$\times 10^{-3}$ & 2.8$\times 10^{-2}$\nl
L = E/$\tau_{dyn}$ (\lsol) & 5.4 & 21.6 \nl
\enddata
\end{deluxetable}

\end{document}